\documentclass[12pt]{article}
\usepackage{cite,amssymb,amsmath}
\usepackage{color}
\usepackage[colorlinks=true,linkcolor=red]{hyperref}
\usepackage[dvipsnames]{xcolor}
\usepackage{cite,amssymb,graphicx,graphics}

\begin{document}

\title{\bf Localization Problem in the 5D Standing Wave Braneworld}

\author{{\bf Merab Gogberashvili}\\
Andronikashvili Institute of Physics,\\
6 Tamarashvili St., Tbilisi 0177, Georgia\\
and \\
Javakhishvili State University, \\
3 Chavchavadze Ave., Tbilisi 0128, Georgia\\
{\sl E-mail: gogber@gmail.com} \\\\
{\bf Pavle Midodashvili}\\
Ilia State University, \\ 3/5 Kakutsa Cholokashvili Ave., Tbilisi 0162, Georgia\\
{\sl E-mail: pmidodashvili@yahoo.com} \\\\
{\bf Levan Midodashvili}\\
Gori Teaching University, \\ 53 Chavchavadze St., Gori 1400, Georgia\\
{\sl E-mail: levmid@hotmail.com}}

\maketitle

\begin{abstract}
We investigate the problem of pure gravitational localization of matter fields within the 5D standing wave braneworld generated by gravity coupled to a phantom-like scalar field. We show that in the case of increasing warp factor there exist normalizable zero modes of spin-0, -1/2, -1, and -2 fields on the brane.
\end{abstract}

\vskip 0.5cm

%%%%%%%%%%%%%%%%%%%%%%%%%%%%%%%%%%%%%%%%%%%%%%%%%%%%%%%%%%%%%%%%%%%

\section{Introduction}

The scenario where our world is associated with a brane embedded in a higher dimensional space-time with non-factorizable geometry\cite{Hi-1,Hi-2,brane-1,brane-2} has attracted a lot of interest with the aim of solving several open questions in modern physics (see\cite{reviews-1,reviews-2,reviews-3,reviews-4} for reviews). Most of examined models were realized as time independent field configurations. However, there have been considered several braneworld models that assumed time-dependent metrics and fields\cite{S-1,S-2,S-3,S-4}. Here we consider non-stationary braneworld scenario recently proposed in\cite{Wave-1,Wave-2,Wave-3}. The braneworld is generated by 5D standing gravitational waves coupled to a phantom-like scalar field in the bulk.

In our model\cite{Wave-1,Wave-2,Wave-3} the bulk standing wave is bounded by the brane and the static part of the gravitational potential, which increases at the extra infinity. So there are some similarities with the experiments of\cite{Nes}, where neutrons were trapped between a reflecting plane at the surface of the Earth and the 'wall' generated by the gravitational potential of the Earth. We have discrete frequencies of standing wave oscillations just as the neutrons had discrete energy levels in those experiments.

Another feature of the model\cite{Wave-1,Wave-2,Wave-3} is the anisotropy of 4D part of the metric. Anisotropic braneworlds\cite{Anistrop-1,Anistrop-2,MSS,NPK,FLSZ,Gh-Ka-1,Gh-Ka-2} have been considered when addressing various issues like anisotropy dissipation during inflation\cite{MSS}, localization of test particles\cite{Gh-Ka-1,Gh-Ka-2}, or braneworld isotropization with the aid of magnetic fields\cite{NPK}. As a general feature it has been established that anisotropic metrics on the brane prevent the bulk from being static\cite{FLSZ,NPK}.

A key requirement for realizing the braneworld idea is that the various bulk fields be localized on the brane. For reasons of economy and avoidance of charge universality obstruction\cite{DuRuTi} one would like to have a universal gravitational trapping mechanism for all fields. However, there are difficulties to realize such mechanism with stationary exponentially warped space-times. In the existing (1+4)-dimensional models spin-$0$ and spin-$2$ fields can be localized on the brane with the decreasing warp factor\cite{brane-1,brane-2}, spin-$1/2$ field can be localized with the increasing warp factor\cite{BaGa}, and spin-$1$ fields are not localized at all\cite{Po}. In the case of 6D models it was found that spin-$0$, spin-$1$ and spin-$2$ fields are localized on the brane with the decreasing warp factor and spin-$1/2$ fields again are localized with the increasing warp factor\cite{Od}. There exist also 6D models with non-exponential warp factors providing gravitational localization of all kinds of bulk fields on the brane\cite{6D-1,6D-2,6D-3,6D-4,6D-5}, however, these models require introduction of unnatural gravitational sources.

In this paper we show that standing wave braneworld model\cite{Wave-1,Wave-2,Wave-3} can provide universal gravitational trapping of zero modes of all kinds of matter fields in the case of rapid oscillations of standing waves in the bulk. To clarify the mechanism of localization  let us remind that standing electromagnetic waves, so-called optical lattices, can provide trapping of various particles by scattering, dipole and quadruple forces\cite{Opt-1,Opt-2,Opt-3}. It is known that the motion of test particles in the field of a gravitational wave is similar to the motion of charged particles in the field of an electromagnetic wave\cite{Ba-Gr}. Thus standing gravitational waves could also provide confinement of matter via quadruple forces. Indeed, the equations of motion of the system of spinless particles in the quadruple approximation has the form\cite{Dix}:
\begin{equation} \label{quad}
\frac{Dp^\mu}{ds}= F^\mu = -\frac 16 J^{\alpha\beta\gamma\delta}D^\mu R_{\alpha\beta\gamma\delta}~,
\end{equation}
where $p^\mu$ is the total momentum of the matter field and $J^{\alpha\beta\gamma\delta}$ is the quadruple moment of the stress-energy tensor for the matter field. The oscillating metric due to gravitational waves should induce a quadruple moment in the matter fields. If the induced quadruple moment is out of phase with the gravitational wave the system energy increases in comparison with the resonant case and the fields/particles will feel a quadruple force, $F^\mu$, which ejects them out of the high curvature region, i.e. it would localize them at the nodes.

We proceed as follows. In Section 2 we recall basic ingredients of the 5D standing wave braneworld model\cite{Wave-1,Wave-2,Wave-3}. In Section 3 we impose boundary conditions to choice exact form of background metric for matter fields and also list the approximations we use. Then, in Sections 4, 5, 6 and 7 we demonstrate existence of zero modes of spin-$0$, -$1$, -$1/2$ and -$2$ particles on the brane, respectively. Short conclusions can be found in Section 8.

%%%%%%%%%%%%%%%%%%%%%%%%%%%%%%%%%%%%%%%%%%%%%%%%%%%%%%%%%%%%%%%%%%%%

\section{Background solution}

5D standing wave braneworld\cite{Wave-1,Wave-2,Wave-3} is generated by gravity coupled to a non-self-interacting scalar phantom-like field (different brane models with phantom fields can be found in\cite{phantom-1,phantom-2,phantom-3,phantom-4,phantom-5,phantom-6}), which depends on time and propagates in the bulk. The action of the model has the form:
\begin{equation} \label{action}
S = \int d^5x \sqrt{g} \left[\frac{1}{16 \pi G_5} \left( R -
2\Lambda_{5}\right) - \frac{1}{2}g^{MN}\partial_M \phi
\partial_N\phi \right]~,
\end{equation}
where capital Latin indexes refer to 5D space-time and $G_5$ and $\Lambda _5$ are 5D Newton and cosmological constants respectively.

To avoid the well-known problems with stability, which occur with ghost fields, we can associate, for example, the ghost-like field $\phi$ with the geometrical scalar field in a 5D integrable Weyl model\cite{Weyl-1,Weyl-2,Weyl-3,Weyl-4,Weyl-5}. In the Weyl model a massless scalar, either an ordinary or ghost-like scalar, appears through the definition of the covariant derivative of the metric tensor:
\begin{equation} \label{D}
D_A g_{MN} = g_{MN}\partial_A \phi ~.
\end{equation}
This is a generalization of the Riemannian case where the covariant derivative of the metric is zero. The gravitational action for the Weyl 5D integrable model can be written as:
\begin{equation}\label{grav-weyl}
S_W = \int d^5x \sqrt{g} \left[\frac{1}{16 \pi G_5} \left( R -
2\Lambda_{5}\right) - (6-5\xi)g^{MN}\partial_M \phi
\partial_N\phi  \right]~,
\end{equation}
where $\xi$ is an arbitrary constant. For $\xi = 11/10$ the action (\ref{grav-weyl}) exactly coincides with (\ref{action}). So we can start with a 5D Weyl model and require that we have Riemann geometry on the brane by assuming that the geometrical scalar field $\phi$ is independent of 4D spatial coordinates and vanishes on the brane.

The definition of $\phi$ via (\ref{D}) avoids the usual instability problems of ghost fields since the geometrical fields have specific couplings with matter fields. They alter only lengths of vectors after parallel transport due to the assumption (\ref{D}), but have not dangerous couplings like $\phi \bar{\psi} \psi$ and it is known that the Weyl model is stable for any value of $\xi$.

Situation is similar to the case with other geometrical structures. For instance, introduction of negative cosmological constants, or non-tensor character of Christoffel symbols does not lead to instability of particle physics models constructed in tangent spaces.

The Einstein equations for the action (\ref{action}) read:
\begin{equation} \label{field-eqns}
R_{MN} - \frac{1}{2} g_{MN} R =  8\pi G_5 T_{MN} - \Lambda _5 g_{MN}~.
\end{equation}
We use the metric {\it ansatz}:
\begin{equation} \label{metricA}
ds^2 = e^{2a|r|}\left( dt^2 - e^{u}dx^2 - e^{u}dy^2 - e^{-2u}dz^2
\right) - dr^2~,
\end{equation}
where $a$ is a constant, which corresponds to brane width. The peculiarity of the model (\ref{metricA}) is that the brane, located at $r = 0$, possesses anisotropic oscillations and sends a wave into the bulk (as in\cite{GMS-1,GMS-2}), i.e. the brane is warped along the spatial coordinates through the factors $\sim e^{u(t,r)}$, which depend on time $t$ and the extra coordinate $r$.

The phantom-like scalar field $\phi (t,r)$ obeys the Klein-Gordon equation on the background space-time (\ref{metricA}),
\begin{equation} \label{phi}
\frac{1}{\sqrt{g}}~\partial_M (\sqrt{g} g^{MN}\partial_N \phi) =
e^{-2a|r|}\ddot \phi - \phi'' - 4 a~ sgn (r) \phi' = 0 ~,
\end{equation}
where $sgn (r)$ is the sign function, overdots and primes mean derivatives with respect to $t$ and $r$ respectively, and determinant for our {\it ansatz} (\ref{metricA}) is equal to
\begin{equation} \label{determinant}
\sqrt g = e^{4a|r|}~.
\end{equation}

We further rewrite the Einstein - scalar field equations in the form:
\begin{equation} \label{field-eqns1}
R_{MN} = - \partial_M \sigma\partial_N
\sigma + \frac{2}{3} g_{MN} \Lambda _5~,
\end{equation}
where the gravitational constant has been absorbed in the definition of the scalar field:
\begin{equation}
\label{sigma} \sigma = \sqrt{8 \pi G_5} ~ \phi ~.
\end{equation}
It turns out that the system (\ref{field-eqns1}) is consistent only when the fields $\sigma$ and $u$ are related with each other in the form \cite{Wave-1,Wave-2,Wave-3}:
\begin{equation} \label{sigma=u}
\sigma (t,r) = \sqrt{\frac{3}{2}}~u(t,r)~.
\end{equation}

The anisotropy of the metric (\ref{metricA}) forces us to define the energy-momentum tensor of the brane with different tensions along different directions:
\begin{eqnarray} \label{tensormixto}
\tau^{M}_{N}= \delta{(r)}~\mbox{\rm
diag}[\lambda_t,\lambda_x,\lambda_y, \lambda_z,0]~, ~~~~~ (\lambda_x
= \lambda_y)~.
\end{eqnarray}
The tensions $\lambda_m$ ($m=t,x,y,z$) in general depend only on time. Taking into account (\ref{tensormixto}) the field equations (\ref{field-eqns1}) change as follows:
\begin{equation} \label{field-eqnsDELTA}
R_{MN} = - \partial_{M}
\sigma\partial_{N} \sigma + \frac{2}{3} g_{MN} \Lambda _5 + 8\pi G_5 \bar{\tau}_{MN}~,
\end{equation}
where the reduced energy-momentum tensor,
\begin{equation}
\bar{\tau}_{MN} = \tau_{MN} - \frac{1}{3}g_{MN}\tau~,
\end{equation}
corresponds to the matter content on the brane, and takes the form:
\begin{eqnarray}\label{reduced}
\bar{\tau}_{MN}=\frac{1}{3}\delta{(r)}~\mbox{\rm
diag}\left[\right.2\lambda_t + 2e^{-u}\lambda_x + e^{2u}\lambda_z , e^{u}\lambda_t + \lambda_x - e^{3u}\lambda_z, \\
e^{u}\lambda_t + \lambda_x - e^{3u}\lambda_z, e^{-2u}\lambda_t -
2e^{-3u}\lambda_x + 2\lambda_z,  0 \left.\right]. \nonumber
\end{eqnarray}

The non-zero components of the Ricci tensor for the metric (\ref{metricA}) read:
\begin{eqnarray} \label{ricci}
R_{tt} &=& e^{2a|r|} \left[-\frac {3}{2} e^{-2a|r|} \dot u ^2 + 4a^2 + 2a\delta(r)\right]~, \nonumber \\
R_{xx}&=&R_{yy}= e^{2a|r|+u}\left[ \frac {1}{2} e^{-2a|r|}\ddot u
- 4a^2 - 2a\epsilon(r)u' - 2a\delta(r) -\frac {1}{2} u'' \right]~, \nonumber \\
R_{zz} &=& e^{2a|r|-2u}\left[ - e^{-2a|r|}\ddot u - 4a^2 + 4a\epsilon(r)u' - 2a\delta(r) + u''\right]~, \\
R_{rr} &=& -\frac 32 u'^2 - 4a^2 - 8a\delta(r)~, \nonumber\\
R_{rt} &=& -\frac 32 \dot uu' ~. \nonumber
\end{eqnarray}
Using the fine tuning:
\begin{equation}\label{L=a}
\Lambda_ 5 = 6 a^2~,
\end{equation}
the system of Einstein equations (\ref{field-eqnsDELTA}) reduces to a single ordinary differential equation,
\begin{equation}\label{eqntou}
e^{-2a|r|}~\ddot u - u'' - 4a~sgn(r) u' = 0~.
\end{equation}
In addition, for the brane tensions $\lambda_m$ we have \cite{mm}:
\begin{eqnarray}
\lambda_t &=& -\frac{3a}{4\pi G_5} \delta \left(r\right)~, \nonumber \\
\lambda_x &=& \lambda_y = \frac{e^{u\left(t,0\right)}}{8\pi G_5} \left(6a-\frac{1}{2}[u']\right)\delta \left(r\right) ~,\\
\lambda_z &=& \frac{e^{-2u\left(t,0\right)}}{8\pi G_5}\left(6a+[u']\right)\delta \left(r\right)~, \nonumber \\
\lambda_r &=& 0~, \nonumber
\end{eqnarray}
where $u\left(t,0\right)$ and $[u']$ denote the value of $u(t,r)$ and the jump of its first derivative at $r=0$, respectively.

The standing wave solution to (\ref{eqntou}) can be constructed by implementing the {\it ansatz}:
\begin{equation} \label{separation}
u(t,r) = C \sin (\omega t) f(r)~,
\end{equation}
where $C$ and $\omega$ are real constants. From (\ref{eqntou}) we get the equation for the radial function:
\begin{equation} \label{f}
f'' + 4a~sgn(r) f' + \omega^2 e^{-2 a |r|}f = 0 ~.
\end{equation}
The general solution to this equation has the following form:
\begin{equation} \label{fsol}
f(r) = e^{-2a|r|} \left[A~ J_2\left( \frac{\omega}{|a|} e^{-a|r|}
\right) +  B~ Y_2\left( \frac{\omega}{|a|} e^{-a|r|}
\right)\right],
\end{equation}
where $A,B$ are arbitrary constants and $J_2$ and $Y_2$ are second-order Bessel functions of the first and second kind, respectively.

%%%%%%%%%%%%%%%%%%%%%%%%%%%%%%%%%%%%%%%%%%%%%%%%%%%%%%%%%%%%%%%%%%%%%%%%%

\section{Boundary conditions and averages}

As pointed out in\cite{Wave-1,Wave-2,Wave-3} the ghost-like field $\sigma (t,r)$, along with the metric oscillations $u(t,r)$, must be unobservable on the brane. Taking into account the relations (\ref{sigma=u}) and (\ref{separation}) we can accomplish this requirement by setting the boundary condition for the function (\ref{fsol}):
\begin{equation}
\left. f (r) \right|_{r=0} = 0~.
\end{equation}
Since $J_2$ and $Y_2$ are oscillatory functions, in the case of increasing ($a>0$)/decreasing ($a<0$) warp factor for some fixed values of the constants $A$ and $B$ the function $f(r)$ can have finite/infinite number of zeros. Thus the above boundary condition can be written in the form which quantizes the oscillation frequency, $\omega$, of the standing wave in terms of the curvature scale $a$, i.e.
\begin{equation} \label{quantize}
\frac{\omega}{|a|} = X_n~,
\end{equation}
where $X_n$ is the $n^{th}$ zero of the function $f(r)$. Correspondingly, the nodes of standing wave, the points where the functions $\sigma (t,r)$ and $u(t,r)$ vanish, can be considered as 4D space-time `islands', where the matter particles are assumed to be bound.

For simplicity in this paper we assume that the constant $A$ in the solution (\ref{fsol}) is zero and the oscillatory metric function (\ref{separation}) has the form:
\begin{equation} \label{u}
u \left( t,r \right) = B ~ \sin \left( \omega t \right) e^{-2a|r|} Y_2\left(\frac {\omega}{|a|}~ e^{-a |r| }\right)~.
\end{equation}

Physical solutions described by Bessel functions usually contain the first kind functions, $J_n$, because of their regularity at the origin. However, in our case we can use even Bessel function of the second kind, $Y_n$, since the argument in (\ref{u}), $e^{-a |r|} \omega / |a|$, is always positive and becomes zero only for the case of increasing warp factor ($a > 0$) at $r \to \pm \infty$.

To show explicitly pure gravitational localization of matter fields on the brane by the metric (\ref{metricA}) we consider the case with the increasing warp factor $a>0$. The oscillatory function (\ref{u}) is zero at the position of the brane, $r = 0$, due to fine tuning (\ref{quantize}), where $X_n$ now is one of the zeros of $Y_2$. In this paper we explore the case when the Bessel function $Y_2$ has a single zero, i.e. we assume that
\begin{equation}\label{FirstZerosY}
\frac{\omega}{a} = X_1 \approx 3.38~.
\end{equation}

In the equations of matter fields the oscillatory function (\ref{u}) enters via some exponential functions:
\begin{equation} \label{e-br}
e^{bu} = \sum \limits_{n = 0}^{+\infty } \frac{\left( bu \right)^n}{n!}~,
\end{equation}
where $b$ is a constant. We suppose that the frequency of standing waves $\omega$ in the oscillatory metric function $u(t,r)$ is much larger than frequencies associated with energies $E$ of particles on the brane, i.e.
\begin{equation}
\omega \gg E~.
\end{equation}
In this case we can perform time averaging of oscillating exponents in the equations of matter fields.

Using the expression:
\begin{equation}
\frac{\omega }{2\pi} \int\limits_0^{2\pi/\omega} \left[ \sin (\omega t)\right]^m dt = \left \{
\begin{array} {lr}
0 & ( m = 2n + 1)\\
\frac {2^{-m}m!}{ \left[ (m/2)! \right]^2} & (m = 2n)
\end{array}
\right.
\end{equation}
for the time averages of oscillating exponents (\ref{e-br}) we get the simple formula \cite{Gr-Ry}:
\begin{equation} \label{e-u}
\left\langle e^{bu} \right\rangle = \sum \limits_{n = 0}^{+\infty } \frac{ Z^{2n}}{2^{2n}\left( n! \right)^2} = I_0(Z) ~,
\end{equation}
where $I_0$ is the modified Bessel function of the zero order of the argument:
\begin{equation} \label{Z}
Z = |bB|e^{- 2a|r|} Y_2\left( \frac{\omega }{a}e^{ - a|r|} \right)~.
\end{equation}
Let us also display some equalities for time averages of oscillatory functions \cite{GMM}:
\begin{equation}\label{AdditionalFacts}
\left\langle u \right\rangle  = \left\langle
u' \right\rangle= \left\langle \frac{\partial
u}{\partial t} \right\rangle  =\left\langle \frac{\partial
u}{\partial t}e^{ - u} \right\rangle  = 0~,
\end{equation}
where the prime denotes the derivative with respect to the extra coordinate $r$.

Now we are ready to consider localization problem of various matter fields on the brane within the standing wave braneworld model with background metric (\ref{metricA}), where the oscillatory function $u(t,r)$ has the form (\ref{u}).

%%%%%%%%%%%%%%%%%%%%%%%%%%%%%%%%%%%%%%%%%%%%%%%%%%%%%%%%%%%%%%%%%%%%%%%%

\section{Localization of scalar fields}

We start with the problem of localization of massless scalar fields defined by the 5D action\cite{GMM}:
\begin{equation} \label{Sphi}
S_\Phi = - \frac 12 \int \sqrt{g} dx^4dr ~ g^{MN}\partial_M\Phi \partial_N\Phi~.
\end{equation}
Corresponding Klein-Gordon equation,
\begin{equation}\label{ScalFieldEqn}
\frac{1}{\sqrt g}~\partial_M \left( \sqrt g g^{MN}\partial_N \Phi \right) = 0~,
\end{equation}
can be written as:
\begin{eqnarray}\label{Equation1}
\left[ \partial_t^2 - e^{- u}( \partial_x^2 + \partial_y^2) - e^{2u} \partial_z^2\right]\Phi = e^{2a|r|}\left( e^{4a|r|} \Phi' \right)'.
\end{eqnarray}
We look for the solution of this equation in the form:
\begin{equation}\label{Solution1}
\Phi \left( {t,x,y,z,r} \right) = \Psi (t,r)\chi (x,y)\xi (z)~.
\end{equation}
This separation of variables transforms (\ref{Equation1}) into the system of the equations,
\begin{eqnarray} \label{system}
\left(\partial _x^2 + \partial _y^2 \right)\chi + \left( p_x^2+p_y^2\right)\chi &=& 0 ~, \nonumber \\
\partial_z^2\xi + p_z^2\xi &=& 0 ~, \\
\partial_t^2\Psi + \left[ (p_x^2 + p_y^2)e^{-u} + p_z^2 e^{2u} \right]\Psi &=& e^{2a|r|}\left( e^{4a|r|}\Psi' \right)'. \nonumber
\end{eqnarray}
On the brane, where $u\approx 0$ and anisotropies are absent, the parameters $p_x, p_y$ and $p_z$ can be regarded as momentum components along the brane. Their exact physical meaning is not clear in general case since (\ref{system}) are equations with variable coefficients.

Now we separate variables in the last equation of the system (\ref{system}),
\begin{equation}
\Psi ( t,r) = e^{iEt} \varsigma (r)~,
\end{equation}
where the energy $E$ for 4D massless scalar modes on the brane obeys the dispersion relation:
\begin{equation}
E^2 = p_x^2 + p_y^2 + p_z^2~.
\end{equation}
As it was mentioned in previous section, when the frequency of standing waves $\omega$ is much larger than frequencies associated with the energies of the particles on the brane,
\begin{equation}
\omega \gg E~,
\end{equation}
we can perform time averaging of oscillatory functions. Then for the extra factor $\varsigma (r)$ of the scalar field wave function we obtain the equation:
\begin{equation}\label{var}
\left( e^{4a|r|}\varsigma' \right)' - e^{2a|r|}K(r) \varsigma  = 0~,
\end{equation}
which contains the time-independent factor:
\begin{equation} \label{K(r)}
K(r) = \left(\left\langle e^{-u} \right\rangle -1\right)\left(
p_x^2 + p_y^2 \right) + \left(\left\langle e^{2u} \right\rangle
-1\right)p_z^2 ~.
\end{equation}

It is more convenient to put (\ref{var}) into the form of an analogue non-relativistic quantum mechanical problem by making the change:
\begin{equation} \label{varsigma}
\varsigma (r) = e^{-2a|r|} \psi_s (r)~.
\end{equation}
For $\psi_s (r)$ we find:
\begin{equation}\label{psi-s}
\psi_s'' - U_s(r) \psi_s = 0~,
\end{equation}
where the function
\begin{equation}\label{U-s}
U_s(r)= 4a\delta (r) + 4a^2 + e^{-2a|r|}K(r)
\end{equation}
is the analog of non-relativistic potential. FIG. 1 shows behaviour of $U_s(r)$ in the case defined by (\ref{FirstZerosY}).

%%%%%%%%%%%%%%%%%%%%%%%%%%%%%%%%%%%%%%%%%%%%%%%%%%%%%%%%%%%%%%%%

\begin{figure}[ht]
\begin{center}
\includegraphics[width=0.5\textwidth]{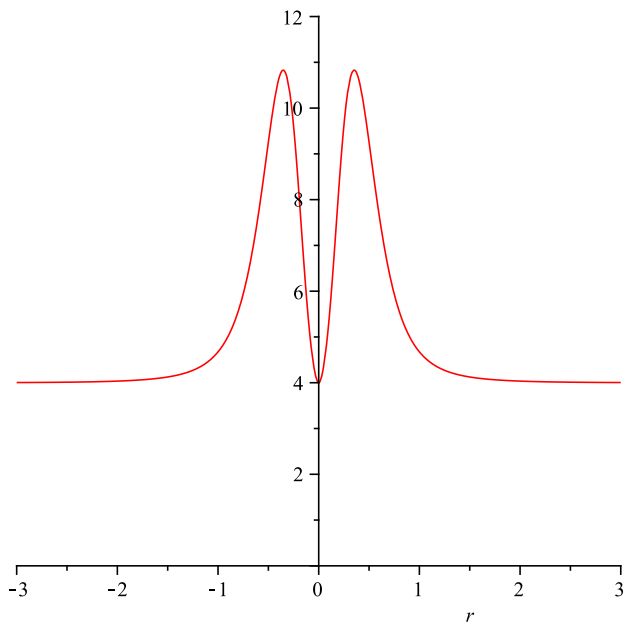}
\caption{The effective bulk potential (\ref{U-s}).}
\end{center}
\end{figure}

%%%%%%%%%%%%%%%%%%%%%%%%%%%%%%%%%%%%%%%%%%%%%%%%%%%%%%%%%%%%%%%%

To study general behaviour of extra part of the scalar field zero mode wave function we explore (\ref{psi-s}) in two limiting regions, far from and close to the brane. The function $K(r)$, which describes oscillatory properties of standing waves, has the following asymptotical forms:
\begin{eqnarray}
\left. K (r)\right|_{r \to 0} &\sim& r^2~, \nonumber \\
\left. K (r)\right|_{r \to \infty} &\sim& const~.
\end{eqnarray}

So close to the brane, $r \to 0$, the equation (\ref{psi-s}) obtains the form:
\begin{equation}\label{phi-s-zero}
\psi_s '' - \left[ 4a \delta (r) + 4a^2 \right]\psi_s = 0~,
\end{equation}
with the solutions
\begin{equation} \label{phi-s-0}
\left. \psi_s (r)\right|_{r \to 0} \sim e^{2a|r|}~.
\end{equation}
At the infinity the equation (\ref{psi-s}) again reduces to (\ref{phi-s-0}) (but without delta-function of course) with the two solutions,
\begin{equation} \label{phi-s-infinity}
\left. \psi_s (r)\right|_{r \to \infty} \sim e^{\pm 2a|r|}~.
\end{equation}

Resemblance of the equations far from and close to the brane is not surprising since standing waves in the bulk are bounded by the brane at $r=0$ and by the gravitational potential at the infinity, i.e. these two regions correspond to the nodes of the waves.

To have normalizable zero mode we impose the boundary conditions:
\begin{eqnarray} \label{varsigma-boundary}
\left. \varsigma ' (r)\right|_{r=0} &=& 0~, \nonumber \\
\left. \varsigma  (r)\right|_{r \to \infty} &=& 0~.
\end{eqnarray}
Then using the definition (\ref{varsigma}) and the solutions (\ref{phi-s-0}) and (\ref{phi-s-infinity}) we find:
\begin{eqnarray} \label{solution-s}
\left. \varsigma (r)\right|_{r \to 0} &\sim& const ~, \nonumber \\
\left. \varsigma (r)\right|_{r \to \infty} &\sim&  e^{-4a|r|}~.
\end{eqnarray}
So $\varsigma (r)$ has a maximum on the brane and falls off at the infinity as $e^{-4a|r|}$. In the action of scalar fields (\ref{Sphi}) the determinant (\ref{determinant}) and the metric tensor with upper indices give the total exponential factor $e^{2a|r|}$, which obviously increases for $a > 0$. This is the reason why in the original brane models\cite{brane-1,brane-2} scalar field zero modes, with the constant $r$-depended extra part of the wave function, can be localized on the brane only in the case of decreasing warp factor (i.e. $a<0$). In our model the extra part of wave function (\ref{solution-s}) is not a constant, moreover, for $a > 0$ it contains the exponentially decreasing factor $e^{-4a|r|}$. For such extra dimension factor the integral over $r$ in the action (\ref{Sphi}) is convergent, i.e. 4D scalar fields are localized on the brane.

%%%%%%%%%%%%%%%%%%%%%%%%%%%%%%%%%%%%%%%%%%%%%%%%%%%%%%%%%%%%%%%%%%%%%%%%

\section{Localization of vector fields}

For simplicity in this paper we investigate only the $U(1)$ vector field\cite{GMM-2}, the generalization to the case of non-Abelian gauge fields is straightforward. The action of vector field is:
\begin{equation}\label{VectorAction}
S_{\rm{v}} = - \frac14\int d^5x\sqrt g~ g^{MN}g^{PR}F_{MP}F_{NR}~,
\end{equation}
where
\begin{equation} \label{F}
F_{MP} = \partial _M A_P - \partial _P A_M
\end{equation}
is the 5D vector field tensor.

Incidentally, in 5D space-time implementation of the pure gravitational trapping mechanism of vector fields on the brane remains most problematic. In the original brane models extra dimension parts of the vector field zero modes are constant functions. So, taking into account that in the action integral (\ref{VectorAction}) the extra coordinate parts of the determinant (\ref{determinant}) and two metric tensors with upper indices cancel each other, the vector field zero modes are not localizable on the brane for any sign  of $a$. Because of this fact there was proposed some non-gravitational trapping mechanisms, for example\cite{DSh}. Here we want to show that the standing wave braneworld metric (\ref{metricA}) provides pure gravitational localization of vector field zero modes on the brane.

The action (\ref{VectorAction}) gives the system of five equations:
\begin{equation}\label{VectorFieldEquations}
\frac{1}{\sqrt g }\partial _M\left( \sqrt g ~g^{MN}g^{PR}F_{NR}
\right) = 0 ~.
\end{equation}

Let us seek for the solution to the system (\ref{VectorFieldEquations}) in the form:
\begin{eqnarray}\label{VectorDecomposition}
A_t(x^C) &=& \upsilon (r)~a_t(x^\nu)~, \nonumber \\
A_x(x^C) &=& e^{u(t,r)} \upsilon  (r)~a_x(x^\nu)~, \nonumber \\
A_y(x^C) &=& e^{u(t,r)} \upsilon  (r)~a_y(x^\nu)~, \\
A_z(x^C) &=& e^{-2u(t,r)} \upsilon  (r)~a_z(x^\nu)~, \nonumber \\
A_r(x^C) &=& 0~,\nonumber
\end{eqnarray}
where $u(t, r)$ is the oscillatory metric function (\ref{u}), $a_{\mu}(x^\nu)$ denote the components of 4D vector potential (Greek letters are used for 4D indices) and scalar factor $\upsilon (r)$ depends only on the extra coordinate $r$. The last expression in (\ref{VectorDecomposition}) is in fact the 5D gauge condition.

We require existence of flat 4D vector waves localized on the brane,
\begin{equation} \label{Factors}
a_\mu\left(x^\nu\right) \sim \varepsilon_\mu e^{i(Et + p_x x + p_y y + p_z z)} ~,
\end{equation}
where $E$, $p_x$, $p_y$, $p_z$ are components of energy-momentum along the brane. Solutions of this kind exist only on the brane, where $u = 0$, and in the case when $\omega$ is much larger than frequencies associated with the energies of the particles on the brane $E$. For such high frequencies of bulk standing waves we can perform time averaging of oscillatory functions in (\ref{VectorFieldEquations}).

Taking into account the equalities (\ref{AdditionalFacts}) time averaging of the fifth equation of the system (\ref{VectorFieldEquations}),
\begin{equation}
\partial_\alpha \left( g^{\alpha\beta}  A_\beta '\right) = 0~,
\end{equation}
gives the Lorenz-like gauge condition:
\begin{equation} \label{GaugeCondition2}
g^{\alpha\beta}\partial_\alpha A_\beta =
\eta^{\alpha\beta}\partial_\alpha a_\beta = 0~,
\end{equation}
where $\eta_{\alpha\beta}$ denotes the metric of 4D Minkowski space-time. The equation (\ref{GaugeCondition2}) together with the last expression of (\ref{VectorDecomposition}) can be considered as the full set of imposed gauge conditions.

Performing time averaging and making use of (\ref{AdditionalFacts}) and (\ref{GaugeCondition2}), remaining four equations of the system (\ref{VectorFieldEquations}),
\begin{equation}\label{}
\partial _\gamma\left( g^{\gamma\delta}g^{\beta\alpha}F_{\delta\alpha} \right) - \frac{1}{\sqrt g }\left( \sqrt g ~g^{\beta\alpha}A'_\alpha \right)' = 0 ~,
\end{equation}
reduce to:
\begin{equation} \label{system-a}
\upsilon ~g^{\alpha \delta}\partial _\alpha\partial _\delta a_\beta + e^{-2a|r|} \left( e^{2a|r|}\upsilon' \right)' a_\beta = 0 ~.
\end{equation}
In the case of the zero mode (\ref{Factors}), the system (\ref{system-a}) gives the single equation for $\upsilon (r)$:
\begin{equation}\label{GeneralEquation}
\left( e^{2a|r|}\upsilon ' \right)' - K(r) \upsilon  = 0~,
\end{equation}
where $K(r)$ is defined by (\ref{K(r)}). By making the change:
\begin{equation}\label{FunctionChange}
\upsilon (r) = e^{-a|r|}\psi_{\rm{v}}(r)~,
\end{equation}
as in the case of scalar field (\ref{psi-s}), we put the equation (\ref{GeneralEquation}) into the form of a non-relativistic quantum mechanical problem:
\begin{equation}\label{psi-v}
\psi_{\rm{v}}''(r) - U_{\rm{v}}(r)\psi_{\rm{v}}(r) = 0~,
\end{equation}
where the potential
\begin{equation}\label{U-v}
U_{\rm{v}}(r)= 2a\delta(r) + a^{2} + e^{-2a|r|} K(r)
\end{equation}
differs from the analogues potential for scalar fields (\ref{U-s}) (see FIG. 1) only in constant factors in the first and second terms. So the solutions to (\ref{psi-v}) close to and far from the brane are:
\begin{eqnarray} \label{}
\left. \psi_{\rm{v}} (r)\right|_{r \to 0} &\sim& e^{a|r|} ~, \nonumber \\
\left. \psi_{\rm{v}} (r)\right|_{r \to \infty} &\sim&  e^{\pm a|r|}~.
\end{eqnarray}
Taking into account the definition (\ref{FunctionChange}) and imposing boundary conditions analogues to (\ref{varsigma-boundary}), for the extra dimension factor of the vector field zero mode wavefunction we get:
\begin{eqnarray} \label{solution-v}
\left. \upsilon (r)\right|_{r \to 0} &\sim& const ~, \nonumber \\
\left. \upsilon(r)\right|_{r \to \infty} &\sim&  e^{-2 a|r|}~.
\end{eqnarray}

It's easy to see that the extra factor $\upsilon (r)$ of the vector field zero mode wave function has maximum on the brane and falls off at the infinity as $e^{-2a|r|}$. For such zero mode solution integrals over $r$ in the action of vector fields (\ref{VectorAction}) are convergent, therefore 4D vector fields are localized on the brane.

Indeed, using the {\it ansatz} (\ref{VectorDecomposition}) and the equalities (\ref{AdditionalFacts}), time-averaged components of the vector field tensor (\ref{F}) can be written in the form:
\begin{eqnarray} \label{FF}
\left\langle F_r^\nu \left(x^C \right)\right\rangle &=& \left\langle g^{\nu\beta}F_{r\beta} \right\rangle = e^{-2a|r|}\upsilon' ~a^\nu \left(x^\alpha \right)~, \nonumber \\
\left\langle F_\nu^\mu \left(x^C \right) \right\rangle &=&
\left\langle g^{\nu\beta}F_{\mu\beta} \right\rangle =
e^{-2a|r|}\upsilon  \left[ f_\nu^\mu \left(x^\alpha \right) +
M_\nu^\mu \left(\left\langle e^{bu} - 1 \right\rangle
\partial_\beta a^\alpha \right)\right]~,
\end{eqnarray}
where
\begin{equation}
f_{\mu\nu} \left(x^\alpha \right) = \partial _\mu a_\nu - \partial _\nu a_\mu
\end{equation}
is the 4D vector field tensor and $M_\nu^\mu $ represents the sum of terms of the type $\left\langle e^{bu} - 1 \right\rangle \partial_\nu a^\mu$. Since the functions $\upsilon '$ and $\left\langle e^{bu} - 1 \right\rangle $ in (\ref{FF}) vanish on the brane, the 5D vector lagrangian at $r = 0$ gets the standard 4D form:
\begin{equation}\label{Lagrangian}
L = - \left. \frac14\sqrt g~ \left\langle F_M^P \right\rangle \left\langle F_P^M \right\rangle \right|_{r=0} = -\frac14 \upsilon ^2(0) ~f_\alpha^\beta  f_\beta^\alpha ~.
\end{equation}
Of course in general the 5D lagrangian of vector fields is more complicate. For instance, it contains the mass term $\left(\upsilon' F_r^\alpha\right)^2 \sim \left(\upsilon'\right)^2 a_\alpha a^\alpha$, i.e. the zero mode vector particles are massless only on the brane and acquire masses in the bulk. This fact can be considered as the alternative mechanism of localization.

%%%%%%%%%%%%%%%%%%%%%%%%%%%%%%%%%%%%%%%%%%%%%%%%%%%%%%%%%%%%%%%%%%%%%%%%

\section{Localization of massless fermions}

In this section we investigate the localization problem for massless fermions within the standing braneworld model\cite{GMM-3}. For Minkowskian $4\times 4$ gamma matrices ($\{ \gamma^\alpha, \gamma^\beta \} = 2\eta^{\alpha\beta}$) we use the Weyl basis,
\begin{eqnarray}\label{MinkowskianGammaMatrices}
\begin{array}{l}
\gamma ^t =~ \left( {\begin{array}{*{20}{c}}
0&I\\
I&0
\end{array}} \right),~~~
{\gamma ^i} = \left( {\begin{array}{*{20}{c}}
0&-\sigma^i\\
\sigma^i&0
\end{array}} \right), \nonumber \\
\gamma ^5 = i \gamma^t\gamma^x\gamma^y\gamma^z = \left( {\begin{array}{*{20}{c}}
I&0\\
0&-I
\end{array}} \right),
\end{array}
\end{eqnarray}
where $I$ and $\sigma^i$ ($i = x,y,z$) denote the standard $2\times2$ unit and Pauli matrices respectively.

Let us recall that four-component columns represent fermions in 5D, and that 5D gamma matrices can be chosen as:
\begin{eqnarray} \label{Gamma}
\Gamma^A = h_{\bar A}^A\Gamma^{\bar A}~,~~~~~~\left\{ \Gamma ^A,\Gamma ^B \right\} = 2g^{AB}
~, \nonumber \\
\Gamma^{\bar A} = \left(\gamma^t,\gamma^x,\gamma^y,\gamma^z,i\gamma^5 \right)~,
\end{eqnarray}
where $\bar A,\bar B, ...$ refer to 5D local Lorentz (tangent) frame. So according to (\ref{Gamma}) the curved-space gamma matrices $\Gamma^A$ are related to Minkowskian ones as:
\begin{eqnarray}\label{GammaMatricesRelation}
\Gamma ^t &=& e^{ - a|r|}~\gamma ^t ~, \nonumber \\
\Gamma ^x &=& e^{ - a|r| - u/2}~ \gamma ^x~, \nonumber \\
\Gamma ^y &=& e^{ - a|r| - u/2}~\gamma ^y~, \\
\Gamma ^z &=&  e^{ - a|r| + u}~ \gamma ^z~, \nonumber \\
\Gamma ^r &=& i\gamma ^5~. \nonumber
\end{eqnarray}

The {\it f\"{u}nfbein} for our metric (\ref{metricA}),
\begin{eqnarray}
h^{\bar A}_A = \left( e^{ a|r|},e^{ a|r|+ u/2},e^{ a|r| + u/2},e^{ a|r|-u},1 \right)~,\nonumber \\
h^{\bar AA} = g^{AB} h_B^{\bar A}, ~~~ h^A_{\bar A} = \eta_{\bar A \bar B} h^{\bar B A}, ~~~ h_{\bar AA} = \eta_{\bar A \bar B} h^{\bar B}_A ,
\end{eqnarray}
is introduced through the conventional definition:
\begin{equation}\label{VielbeinDefinition}
g_{AB}=\eta _{\bar A\bar B}h^{\bar A}_A h^{\bar B}_B~.
\end{equation}

The 5D Dirac action for free massless fermions can be written as:
\begin{equation}\label{SpinorAction}
S = \int d^5x \sqrt g ~i\overline \Psi \left(x^A\right) \Gamma ^MD_M \Psi \left(x^A\right)~,
\end{equation}
where covariant derivatives are defined as follows:
\begin{equation}
D_A = \partial_A + \frac 14 \Omega_A^{\bar B \bar C} \Gamma_{\bar B} \Gamma_{\bar C}~.
\end{equation}
In the last expression $\Omega_M^{\bar M \bar N}$ denotes the spin-connection:
\begin{eqnarray}\label{Spin-Connection}
\Omega_M^{\bar M \bar N} = - \Omega_M^{\bar N \bar M} = \frac12
\left[ h^{N\bar M}\left( \partial _M h_N^{\bar N} - \partial_N
h_M^{\bar N} \right) - h^{N\bar N}\left( \partial_M h_N^{\bar M} - \partial _N h_M^{\bar M} \right) - \right.  \\
\left. - h_M^{\bar A} h^{P\bar M}h^{Q\bar N} \left( \partial_P h_{Q\bar A} - \partial_Q h_{P\bar A} \right) \right] ~. \nonumber
\end{eqnarray}
The non-vanishing components of the spin-connection in the background (\ref{metricA}) are:
\begin{eqnarray}\label{Spin-ConnectionComponents}
\Omega_t^{\bar t \bar r} &=& - \left( e^{a|r|} \right)'~, \nonumber \\
\Omega_x^{\bar x \bar r} &=& \Omega_y^{\bar y \bar r} = - \left( e^{a|r| + u/2 } \right)'~, \nonumber \\
\Omega_z^{\bar z \bar r} &=& - \left( e^{a|r| - u} \right)' ~,\\
\Omega_x^{\bar x \bar t} &=& \Omega_y^{\bar y \bar t} = \frac{\partial \left( e^{u/2}\right)}{\partial t}~, \nonumber \\
\Omega_z^{\bar z \bar t} &=& \frac{\partial \left(  e^{ - u}\right)}{\partial t}~. \nonumber
\end{eqnarray}

The corresponding to (\ref{SpinorAction}) 5D Dirac equation reads:
\begin{equation}\label{SpinorEquation}
i\Gamma^AD_A\Psi = i\left( \Gamma ^\mu D_\mu + \Gamma ^rD_r \right)\Psi = 0~.
\end{equation}
For the bulk fermion field wave function we use the chiral decomposition:
\begin{equation}\label{Psi}
\Psi \left(x^\nu,r\right) = \psi_L \left(x^\nu\right) \lambda (r) + \psi_R \left(x^\nu\right) \rho (r)~,
\end{equation}
where $\lambda(r)$ and $\rho(r)$ are extra dimension factors of the left and right fermion wave functions respectively. We assume that 4D left and right Dirac spinors,
\begin{equation} \label{chiral}
\gamma^5 \psi_L = - \psi_L~, ~~~~~ \gamma^5 \psi_R = + \psi_R ~,
\end{equation}
correspond to zero mode wave functions, i.e. they satisfy free Dirac equations:
\begin{equation} \label{Dirac-free}
i\gamma^\mu \partial_\mu \psi_L = i\gamma^\mu \partial_\mu \psi_R = 0~.
\end{equation}

Apart from the massless states $\psi_L$ and $\psi_R$, the 5D Dirac equation also have solutions corresponding to massive fermions. In the single brane models the masses of the bounded massive states are typically of order of the energy scale $a$, characterizing the brane as a topological defect in higher-dimensional space-time. These states are very heavy and we do not consider them here.

The solutions of (\ref{Dirac-free}) in our representation can be written in the form:
\begin{eqnarray} \label{psi-free}
\psi_R \left(x^\nu\right) =
\left( \begin{array}{c}R\\0\end{array} \right)
e^{ - i( Et - p_xx - p_yy - p_zz )}~,\nonumber \\
\psi_L \left(x^\nu\right) =
\left( \begin{array}{c}0\\L\end{array} \right)
e^{ - i( Et - p_xx - p_yy - p_zz )}~,
\end{eqnarray}
where the constant 2-spinors $L$ and $R$ satisfy the relations:
\begin{equation} \label{relations}
\left(E + \sigma^ip_i\right)L = \left(E - \sigma^ip_i\right)R = 0~.
\end{equation}

When the frequency $\omega$ of standing waves in the background metric (\ref{metricA}) is much larger than frequencies associated with the energies $E$ of the fermions on the brane we can perform time averaging of the oscillatory functions in the Dirac equation (\ref{SpinorEquation}). Time averages of the Dirac operators are:
\begin{eqnarray}\label{TimeAveragesOfDiracOperators}
\left\langle i\Gamma^tD_t \right\rangle &=& ie^{ -a|r|}~\gamma ^t\partial_t - \frac 12 a ~sgn(r)\gamma^5, \nonumber \\
\left\langle i\Gamma^xD_x \right\rangle &=& ie^{ -a|r|}\left\langle e^{ -u/2} \right\rangle \gamma ^x\partial_x - \frac 12 a ~sgn(r)\gamma^5, \nonumber \\
\left\langle i\Gamma^yD_y \right\rangle &=& ie^{ - a|r|}\left\langle e^{ -u/2} \right\rangle \gamma ^y\partial_y - \frac 12 a ~sgn(r)\gamma^5, \nonumber \\
\left\langle i\Gamma^zD_z \right\rangle &=& ie^{- a|r|}\left\langle e^{ u} \right\rangle \gamma ^z\partial_z - \frac 12 a ~sgn(r)\gamma^5, \\
\left\langle i\Gamma^rD_r \right\rangle &=& -\gamma ^5\partial_r, \nonumber
\end{eqnarray}
and the equation (\ref{SpinorEquation}) can be written in the form:
\begin{eqnarray}\label{SpinorEquation1}
i\left[{\gamma^t}{\partial _t} + \left\langle e^{ u/2} \right\rangle \left(\gamma ^x\partial _x + \gamma ^y\partial _y\right)  + \left\langle e^{-u} \right\rangle \gamma ^z\partial _z\right]\Psi = \nonumber \\
= e^{a|r|}\gamma^5 \left[2a ~sgn(r) + \partial_r \right]\Psi~.~~~
\end{eqnarray}
Using the solutions of free equations (\ref{psi-free}) and the relations (\ref{relations}), the equation (\ref{SpinorEquation1}) can be rewritten as the system:
\begin{equation}\label{L-R}
\left( \begin{array}{*{20}{c}}
-e^{a|r|}\left[2a ~sgn(r) + \partial_r \right]&\sigma^iP_i(r)\\
- \sigma^iP_i(r)&e^{a|r|}\left[2a ~sgn(r) + \partial_r \right]
\end{array} \right) \left( \begin{array}{*{10}{c}} \rho (r)R\\
\lambda (r)L\end{array} \right) = 0~.
\end{equation}
Here we have introduced {\it '$r$-dependent momentum'} $P_i (r)$:
\begin{eqnarray} \label{P-i}
P_x (r) &=& \left(\left\langle e^{-u/2} \right\rangle -1\right) p_x = \left[I_0\left(Z(r)/2\right)-1\right] p_x, \nonumber \\
P_y (r) &=& \left(\left\langle e^{-u/2} \right\rangle - 1 \right)p_y = \left[I_0\left(Z(r)/2\right)-1\right] p_y, \nonumber \\
P_z (r) &=& \left(\left\langle e^{u} \right\rangle - 1 \right)p_z = \left[I_0\left(Z(r)\right)-1\right] p_z, \\
P^2(r) &=& P_x^2 + P_y^2 + P_z^2~.\nonumber
\end{eqnarray}
where $Z(r)$ is defined by (\ref{Z}).

From the second equation of the system (\ref{L-R}) it is straightforward to find
\begin{equation} \label{rho=lambda}
\rho (r)R = e^{a|r|} \frac {\sigma^iP_i(r)}{P^2(r)}\left[2a ~sgn(r) + \partial_r \right]\lambda (r)L~.
\end{equation}
Inserting (\ref{rho=lambda}) into the first equation of (\ref{L-R}) and multiplying the result by $\sigma^iP_i$, we receive the second order differential equation for the function $\lambda (r)$:
\begin{equation}\label{L-Equation}
\lambda'' + \left[ 5a~ sgn( r ) - \frac{P'}{P}\right] \lambda' +
\left[4a \delta (r) + 6a^2 - 2a~sgn(r)\frac{P'}{P} - P^2e^{ -
2a|r|} \right] \lambda = 0~.
\end{equation}
Now, as in the cases of scalar and vector fields, let us investigate this equation far from and close to the brane.

Close to the brane, $r \to \pm 0 $, the {\it '$r$-dependent momentum'} (\ref{P-i}) behaves as:
\begin{equation}
\left. P (r)\right|_{r\to \pm 0} = A r^2 + O(r^3)~,
\end{equation}
where $A$ is constant, and the equation (\ref{L-Equation}) takes the following asymptotic form:
\begin{equation}
\lambda'' + \left[ 5a~ sgn( r ) - \frac{2}{r}\right] \lambda' +
\left[4a \delta (r) + 6a^2 - \frac {4a}{r}~sgn(r) \right] \lambda
= 0~.
\end{equation}
This equation has the unique nontrivial solution:
\begin{equation}\label{L-0}
\lambda (r)|_{r \to \pm 0} = C e^{ - 2a|r|}~,
\end{equation}
where $C$ is a constant.

As it follows from (\ref{rho=lambda}) and (\ref{L-0}), in our setup the right fermionic modes are absent on the brane:
\begin{equation}\label{rho-0}
\rho (r)|_{r \to \pm 0} = 0 ~.
\end{equation}
Such different behavior of the left and right massless fermions on the brane is not surprising, since in our model the effective mass term in (\ref{SpinorEquation1}) is of $\gamma^5$-type,
\begin{equation}
m(r) = 2a \gamma^5sgn(r) e^{a|r|}~,
\end{equation}
with the gap:
\begin{equation}
|m(r) - m(-r)| = 4a \gamma^5 e^{a|r|}~.
\end{equation}

In the second limiting region $r \to \pm \infty $ the function $P'/P$ vanishes and the equation (\ref{L-Equation}) obtains the asymptotic form:
\begin{equation}\label{L-infinity_0}
\lambda'' + 5a~ sgn( r ) \lambda' + 6a^2 \lambda = 0~,
\end{equation}
with the solution:
\begin{equation}\label{L-infinity}
\lambda (r)|_{r \to \pm \infty } \sim e^{ - 3a|r|}~.
\end{equation}
Using (\ref{L-infinity}) from the relation (\ref{rho=lambda}) we find the asymptotic behavior of the extra factor of the right fermion wave function:
\begin{equation}\label{R-infinity}
\rho (r)|_{r \to \pm \infty } \sim e^{ -2 a|r|}~.
\end{equation}

So in our model the extra dimension part $\lambda (r)$ of the bulk left spinor wave function (\ref{Psi}) has the maximum at the origin,
\begin{equation}
\lambda (r)|_{r=0} = C~,
\end{equation}
falls off from the brane, and turns into the asymptotic form (\ref{L-infinity}) at the infinity. When $r\to \infty$ the determinant (\ref{determinant}) in the action integral for 5D fermions (\ref{SpinorAction}) increases as $e^{4a|r|}$ for $a>0$. However, extra dimension factor of left fermions, according to (\ref{L-infinity}), contribute $e^{-6a|r|}$ and overall $r$-depended part in (\ref{SpinorAction}) decreases as $e^{-2a|r|}$. So in the case of left fermions the integral over $r$ in (\ref{SpinorAction}) is convergent, i.e. zero modes of left fermions are localized on the brane.

At the same time, according to (\ref{rho-0}) and (\ref{R-infinity}), right fermion zero modes does not exist on the brane and the extra dimension part $\rho (r)$ of their wave functions at the infinity decreases as $e^{ -2a|r|}$. Therefore for right fermions integral over $r$ in (\ref{SpinorAction}) diverges and zero mode wavefunctions of right fermions actually are not normalizable.

%%%%%%%%%%%%%%%%%%%%%%%%%%%%%%%%%%%%%%%%%%%%%%%%%%%%%%%%%%%%%%%%%

\section{Localization of gravitons}

Finally, let us deal with spin$-2$ graviton. We shall consider the metric fluctuations:
\begin{equation} \label{metric}
ds^2 = e^{2a|r|}\left( g_{\mu\nu} + h_{\mu\nu}\right)dx^\mu dx^\nu - dr^2~,
\end{equation}
where $g_{\mu\nu}$ is the metric tensor of the 4D part of (\ref{metricA}):
\begin{equation} \label{metric-4}
g_{\mu\nu} = \left( 1, - e^{u},  - e^{u},  - e^{-2u}\right) ~.
\end{equation}

Let us suppose that the frequency of standing waves in the bulk $\omega$ is much larger than frequencies of gravitational waves on the brane, i.e. we can replace the oscillatory functions $e^{u(t,r)}$ in (\ref{metric-4}) by their time averages (\ref{e-u}). So, since the brane is placed in a node of the standing wave, close to $r=0$ we can use the approximation:
\begin{equation}
\left\langle e^{u(t,r)} \right\rangle \approx 1 + \left\langle u \right\rangle~.
\end{equation}
This means that the functions $\left\langle u \right\rangle$ can be regarded as r-depended additive terms of $h_{\mu\nu}$ in flat Minkowski space. Then the equations of motion for the fluctuations $h_{\mu\nu}$ are found to be
\begin{equation}\label{graviton}
\frac{1}{\sqrt g}~\partial_M \left( \sqrt g g^{MN}\partial_N h_{\mu\nu} \right) = 0~.
\end{equation}
It turns out that these equations are equivalent to the equation of motion of a spin-$0$ scalar field (\ref{ScalFieldEqn}) if we replace $\Phi$ with $h_{\mu\nu}$. Accordingly, the localization problems for spin-$2$ graviton field and spin-$0$ scalar field, considered in the Section 4, are similar, and gravitons are also localized on the brane.

%%%%%%%%%%%%%%%%%%%%%%%%%%%%%%%%%%%%%%%%%%%%%%%%%%%%%%%%%%%%%%%%

\section{Conclusion}

In this letter we have investigated the localization problem of all kinds of fields (scalar, vector, spinor and tensor) within non-stationary braneworld scenario\cite{Wave-1,Wave-2,Wave-3}, where the braneworld is generated by 5D standing gravitational waves coupled to a phantom-like scalar field in the bulk. It is noteworthy that the trapping of fields on the brane in our model has universal and purely gravitational nature. The conspicuous feature of our model, as opposed to earlier static approaches with decreasing warp factors\cite{brane-1,brane-2}, is the use of the metric \emph{ansatz} with increasing warp factor. By explicit calculations we have shown that scalar, vector and tensor field zero modes are localized on the brane. In the case of fermion fields we have found that the left fermion zero mode is localized on the brane, while the right fermion zero mode does not exist on the brane and its wavefunction actually is not normalizable.

In our opinion one can use the restriction (\ref{FirstZerosY}), which controls number of nodes of the bulk standing waves in our model, i.e. number of parallel 4D space-time islands, to address the old problem - the nature of flavor. In this regard our scenario has to be studied in detail using numerical solutions to the equations of matter fields in the case of interacting fields. Corresponding investigations are currently in progress.

%%%%%%%%%%%%%%%%%%%%%%%%%%%%%%%%%%%%%%%%%%%%%%%%%%%%%%%%%%%%%%%%

\section*{Acknowledgments}

This research was supported by the grant of Shota Rustaveli National Science Foundation $\#{\rm GNSF/ST}09\_798\_4-100$.

%%%%%%%%%%%%%%%%%%%%%%%%%%%%%%%%%%%%%%%%%%%%%%%%%%%%%%%%%%%%%%%%%%

\end{document}